\documentclass[11pt]{article}
\usepackage[textwidth=15.2cm,textheight=22cm]{geometry}
\usepackage{amsmath,amssymb}
\usepackage{latexsym}
\usepackage{multicol}
\usepackage{graphicx}
\usepackage{bm}
\usepackage{tikz}
\usetikzlibrary{shapes.geometric, arrows}
\usepackage{cite}
\usepackage{tikz}
\usetikzlibrary{trees}
\usepackage{varwidth}
\usepackage{pifont}
\usepackage{cancel}
\usepackage{centernot}
\usepackage{mathtools}
\usepackage[utf8]{inputenc}
\usepackage{longtable}
\usepackage{float}
\usepackage{simpler-wick}
\usepackage{xcolor}
\usepackage{dynkin-diagrams}
\newcommand{\xdownarrow}[1]{%
	{\left\downarrow\vbox to #1{}\right.\kern-\nulldelimiterspace}
}
\usetikzlibrary{shapes.geometric, arrows, shadows}

\tikzstyle{forces} = [rectangle, rounded corners, minimum width=2cm, minimum height=0.8cm,text centered, draw=black]
\tikzstyle{spin} = [rectangle, rounded corners, minimum width=1.3cm, minimum height=0.8cm,text centered, draw=black]
\tikzstyle{theory} = [rectangle, rounded corners, minimum width=2cm, minimum height=0.8cm,text centered, draw=gray]
\tikzstyle{arrow} = [thick,->,>=stealth]
\tikzstyle{arrow1} = [thick,->,>=stealth, color=red]
\tikzstyle{line} = [draw, -latex']

\allowdisplaybreaks[1]

\newcommand{\del}{\partial}
\newcommand{\be}{\begin{equation}}
	\newcommand{\ee}{\end{equation}}
\newcommand{\ba}{\begin{eqnarray}}
	\newcommand{\ea}{\end{eqnarray}}

\newcommand{\rom}[1]{\uppercase\expandafter{\romannumeral #1\relax}}
\newcommand{\nbar}[1]{\overline{#1}}

\def\ba{\bar A}

\def\beq{\begin{equation}}
	\def\eeq{\end{equation}}

\newcommand{\nn}{\nonumber}

\newcommand{\ndt}{\noindent}

\def\bea{\begin{eqnarray}}
	\def\eea{\end{eqnarray}}
\def\beas{\begin{eqnarray*}}
	\def\eeas{\end{eqnarray*}}
\def\sla{\raise.15ex\hbox{$/$}\kern-.57em}

\def\parp{\partial^+}

\def\spa#1.#2{\left\langle#1\,#2\right\rangle}
\def\spb#1.#2{\left[#1\,#2\right]}

\date{}
\begin{document}
	\begin{titlepage}
		\begin{flushright}    
			{\small $\,$}
		\end{flushright}
		\vskip 1cm
		\vskip 1cm
		\centerline{\Large{\bf{ An `exceptional' form for symmetries in maximal supergravity}}}
		\vskip 0.3 cm
		\centerline{Sudarshan Ananth and Nipun Bhave}
		\vskip 0.3cm
		\centerline{\it {Indian Institute of Science Education and Research}}
		\centerline{\it {Pune 411008, India}}
		\vskip 1.5cm
		\centerline{\bf {Abstract}}
		\vskip .4cm
		\ndt The global symmetries in maximally supersymmetric theories of gravity in $d\ge4$ are shown to have a universal form in light-cone superspace. The procedure for deriving an all order expression for the $d=4$ case is also discussed.
		\vskip .5cm 
		\vfill
	\end{titlepage}

	\section{Introduction}
	
	\ndt When eleven-dimensional supergravity is dimensionally reduced, it produces maximally supersymmetric theories of gravity in lower dimensions with enhanced global symmetries~\cite{Cremmer:1978ds,Samtleben:2023nwk,Hull:2025jpv,Cremmer:1979up,Cremmer:1978km,Julia:2025rzr}. The relevant symmetry groups transition from classical to exceptional Lie groups and further to Kac-Moody groups in lower dimensions spanning a wide range of algebraic structures~\cite{Julia:1982gx,Nicolai:2024hqh}. 
\\
	
	 \ndt In the light-cone approach to supergravity, where one works exclusively with the physical degrees of freedom, exceptional symmetries are crucial to deriving the interaction vertices in the Lagrangian, order by order in perturbation theory~\cite{Bengtsson:1983pg,Brink:2008qc,Brink:2008hv,Ananth:2024liy}. The exceptional symmetry generators relevant to four and five dimensions, $E_7$ \& $E_6$,  share a very similar form in light-cone superspace. This raises the question of whether classical symmetries in higher dimensional supergravity theories also share the same structure? We will answer this in the affirmative and show that both exceptional and classical symmetries exhibit a `universal form'. To write down such a form for the symmetries, we will identify appropriate superspace variables determined by a combination of maximal supersymmetry (R-symmetry) and the little group in each dimension. \\
	 
	\ndt 	Another motivation to formulate exceptional symmetries in this specific language arises from our interest in the ultra-violet properties of these theories. Light-cone superspace was originally setup to analyze the high energy behavior of supersymmetric gauge theories~\cite{Mandelstam:1982cb,Brink:1982wv}. This framework is ideally suited for a step-by-step analysis of the large-momentum behavior of supergaphs to all orders~\cite{Ananth:2006ac}. Given the close links between Yang-Mills theory and gravity~\cite{Kawai:1985xq,Ananth:2007zy,Bern:1998sv}, particularly the maximally superymmetric versions~\cite{Bern:1998ug, Bern:2010ue}, it seems natural to apply these techniques to supergravity as well. However, doing so requires an intricate understanding of exceptional symmetries - in this formalism - and specifically their effect on supergraphs.  These symmetries are likely to determine which supergraphs are `forbidden' and importantly, which supergraphs `talk' to one another. A universal form in this context, would allow for divergence analysis of supergraphs in diverse dimensions.\\
	 
\ndt We begin with a brief overview of light-cone superspace, focusing on the four dimensional $N=8$ supergravity theory, the most extensively studied of these theories.  \\
	
		\section{A review of $d=4$ light-cone superspace}
	
	\ndt The $\mathcal{N}=8$ light-cone superspace in four-dimensions involves the bosonic coordinates
	\bea
	{x^{\pm}}=\frac{1}{\sqrt 2}\,(\,{x^0}\,{\pm}\,{x^3}\,)\;;\;\; \quad x =\frac{1}{\sqrt 2}\,(\,x^1\,+\,i\,x^2\,)\;;\;\; \quad {\bar x}=(x)^*\ ,
	\eea
	\ndt with the corresponding space-time derivatives being $\del_+,\del_-,\bar{\del},\del$ and the fermionic coordinates ($\theta^m,\bar{\theta}_m$) which transform as the $\mathbf{8}$ and $\mathbf{\bar{8}}$ of $SU(8)$~\cite{Bengtsson:1983pg}. \\

	\ndt The kinematic supersymmetries $q^m,\,\bar{q}_m$ (generators of translations along the fermionic coordinates) are given by
	\be
	\label{su}
	q^{\,m}=-\frac{\del}{\del\bar{\theta}_m}\,+\,\frac{i}{\sqrt 2}\,{\theta^m}\,{\partial^+}\ ;\qquad{{\bar q}_{m}}=\;\;\;\frac{\del}{\del\theta^m}\,-\,\frac{i}{\sqrt 2}\,{{\bar \theta}_m}\,{\partial^+}\,.
	\ee
	
	 \ndt The superspace covariant derivatives $d^m,\,\bar{d}_m$ read
	 \bea
	 \label{du}
	 d^m=-\frac{\del}{\del\bar{\theta}_m}-\frac{i}{\sqrt{2}}\theta^m\del^+\;\;;\;\;\;\;\;\;\bar{d}_m=\frac{\del}{\del\theta^m}+\frac{i}{\sqrt{2}}\bar{\theta}_m\del^+\,.
	 \eea
	 \ndt The kinematic supercharges and the covariant derivatives satisfy the algebra
	 \bea
	 \label{su88}
	 \{q^m,\bar{q}_n\}=i\sqrt{2}\,\delta^m_n\del^+\qquad;\qquad \{q^m,q^n\}=\{\bar{q}_m,\bar{q}_n\}=0\,,\nn\\
	 \{d^m,\bar{d}_n\}=-i\sqrt{2}\,\delta^m_n\del^+\qquad;\qquad \{d^m,d^n\}=\{\bar{d}_m,\bar{d}_n\}=0\,,\nn\\
	 \{q^m,d^n\}=\{\bar{q}_m,\bar{d}_n\}=0\qquad;\qquad\{q^m,\bar{d}_n\}=\{\bar{q}_m,d^n\}=0\,.
	 \eea
	 
	 \ndt The maximal supergravity theory has a total of 256 degrees of freedom. The 128 bosonic degrees of freedom are comprised of the gravitons of helicity $+2$ \& $-2$, twenty-eight gauge fields of helicity $+1$ (and their conjugates) and seventy scalar fields. The 128 fermionic degrees involve the eight gravitinos of helicity $+3/2$, fifty-six gauginos of helicity $+1/2$ and their complex conjugates. The chiral superfield $\phi$ which captures all the degrees of freedom is given by
	\bea
	\label{superfield}
	\begin{split}
		\phi\,(\,y\,)&~=~\frac{1}{{\parp}^2}\,h\,(y)\,+\,i\,\theta^m\,\frac{1}{{\parp}^2}\,{\bar \psi}_m\,(y)\,+\,\frac{i}{2}\,\theta^m\,\theta^n\,\frac{1}{\parp}\,{\bar A}_{mn}\,(y)\ , \\
		\;&-\,\frac{1}{3!}\,\theta^m\,\theta^n\,\theta^p\,\frac{1}{\parp}\,{\bar \chi}_{mnp}\,(y)\,-\,\frac{1}{4!}\,\theta^m\,\theta^n\,\theta^p\,\theta^q\,{\bar D}_{mnpq}\,(y)\ , \\
		\;&+\,\frac{i}{5!}\,\theta^m\,\theta^n\,\theta^p\,\theta^q\,\theta^r\,\epsilon_{mnpqrstu}\,\chi^{stu}\,(y)\ ,\\
		\;&+\,\frac{i}{6!}\,\theta^m\,\theta^n\,\theta^p\,\theta^q\,\theta^r\,\theta^s\,\epsilon_{mnpqrstu}\,\parp\,A^{tu}\,(y)\ ,\\
		\,&+\,\frac{1}{7!}\,\theta^m\,\theta^n\,\theta^p\,\theta^q\,\theta^r\,\theta^s\,\theta^t\,\epsilon_{mnpqrstu}\,\parp\,\psi^u\,(y)\ ,\\
		\,&+\,\frac{4}{8!}\,\theta^m\,\theta^n\,\theta^p\,\theta^q\,\theta^r\,\theta^s\,\theta^t\,\theta^u\,\epsilon_{mnpqrstu}\,{\parp}^2\,{\bar h}\,(y)\ ,
	\end{split}
	\eea
	 
		\ndt with $y=(x^+,x^--\frac{i}{\sqrt{2}}\theta^m\bar{\theta}_m,x,\bar{x})$\footnote{The operation $\frac{1}{\del^+}$ is defined using the Mandelstam prescription~\cite{Mandelstam:1982cb}.}. The superfield satisfies $d^m\,\phi\,=\,0$ (chirality). This is a crucial constraint which ensures that $\phi$ is an irreducible representation of the superPoincar\'e algebra. The superfield also satisfies the inside-out constraint
	\be
	\bar d_{m}^{}\,\bar d_{n}^{}\,\bar d_{p}^{}\,\bar d_{q}^{}\,\phi~=~\frac{1}{ 2}\,\epsilon_{{m}{n}{p}{q}{r}{s}{t}{u}}^{}\,d^{r}_{}\,d^{s}_{}\,d^{t}_{}\,d^{u}_{}\,\bar\phi\ \,,
	\ee 
	 \ndt which is a consequence of maximal supersymmetry.
\vskip 0.3cm

	 \ndt The Lagrangian of the theory is given by
	 \bea
	 \label{La}
	 {\cal L}&=&-\bar\phi\,\frac{\Box}{\partial^{+4}}\,\phi-\;2\,{\kappa}\;{\int}\;{\frac {1}{{\parp}^2}}\;{\nbar \phi}\;\;{\bar \partial}\,{\phi}\;{\bar \partial}\,{\phi}+c.c. \,+\,\mathcal{O}(\kappa^2)\ ,
	 \eea
	 \ndt where $\kappa=\sqrt{8\,\pi\,G}$. This compact expression describes all cubic interactions of component fields (which may run into hundreds of terms in component form). The generators of symmetries in this formalism fall in two categories: kinematical (linearly realized) and dynamical (non-linearly realized). For a detailed discussion on the generators of the superPoincar\'e algebra we refer the reader to appendix A.\\
	 
	\ndt This construction generalizes nicely to higher dimensions $d>4$, the focus of the next section.\\

	\section{A `universal' form for global symmetries}
	
	\ndt The Lorentz group of the eleven-dimensional supergravity theory, upon dimensional reduction to $d$ space-time dimensions, splits into $SO(d-1,1)$, the new space-time Lorentz group, and $SO(11-d)$, the internal symmetry. In many cases, this internal symmetry may be enhanced via duality transformations to a larger group $E_{n(n)}$. In the covariant formulation, this symmetry acts only on the vectors and scalars of the theory. In the light-cone approach, the process of elimination of unphysical modes, mixes all the fields, requiring the entire supermultiplet to transform~\cite{Brink:2008qc} . \\

	\ndt Each global symmetry $G=E_{n(n)}$ has two parts: the maximal compact subgroup $H$ which serves as the R-symmetry group and $K$ (G/H), the coset space. Schematically, the Lie algebra of $G$ is
	\bea
	\label{a3}
	[H,H]=H\;\;,\;\;[H,K]=K\;\;,\;\;[K,K]=H\,.
	\eea
	
	\ndt In order to write an expression for the generators of this global symmetry, we introduce the Grassmann variables $\theta^{m\alpha}$  where $m$ is the R-symmetry index and $\alpha$ is the index corresponding to the spinor representation of the little group. The corresponding kinematic supercharges and covariant derivatives are denoted by $q^{m\alpha}$, $d^{m\alpha}$. These are related to the conjugate representations $\bar{q}_{m\alpha}$, $\bar{d}_{m\alpha}$ by the action of the invariant tensors of the R-symmetry and the little group\footnote{$\,$This is true for $d\,>\,4$.}. It is therefore possible to work with specific components of the supercharges and covariant derivatives. These independent components - as we will show using specific examples - satisfy the relations (\ref{su88}).\\
	
	\ndt In higher dimensions, the same superfield expression as in (\ref {superfield}) may be used, with an added dependence on the extra coordinates, as the degrees of freedom are simply reshuffled. We explain this in detail in appendix B. This superfield is always annhilated by the independent components of $d^{m\alpha}$ in each dimension. It is also important that symmetry transformations acting on the superfield preserve chirality to ensure that the transformed superfield remains an irreducible representation of the superPoincar{\' e} algebra. Further, the generators of the symmetry $G$ must preserve the helicity of the superfield ($G$ being an internal symmetry, commutes with the space-time symmetry). \\
	
	 \ndt The symmetry $H$ is manifest by construction and therefore linearly realized on the superfield~\cite{Brink:2008qc}
	\bea
	\label{l}
	\delta_{H}\phi~=~\omega^m_{\,\,\,n}\, \frac{q^{n\alpha}\,\bar{q}_{m\alpha}}{i\sqrt{2}\,\del^+}\,\phi\,.
	\eea
	\ndt The $\omega^m_{\,\,\,n}$ are constant parameters with appropriate symmetry properties corresponding to the adjoint representation of $H$.\\
	
	\ndt At the lowest order, the coset transformations $K$ shift the scalars by a constant~\cite{Cremmer:1979up,Brink:2008qc}. The scalars in the superfield always appear with four powers of $\theta$ (indices suppressed). By dimension and helicity analysis, an ansatz for $K$ would take the form
	\bea
	\delta_K\phi=\frac{1}{\kappa}\,a_1\,+\,a_2\,+\,\kappa\,a_3+\,\kappa^2\,a_4+...\,\,,
	\eea
	with $a_1\sim\theta^{\,k\alpha}\,\theta^{\,l\beta}\,\theta^{\,m\gamma}\,\theta^{\,n\delta}\,\Sigma_{klmn}$, where $\Sigma_{klmn}$ are constant parameters with appropriate symmetry properties. The commutators (\ref{a3}) and the fact that $H$ is linear, imply that $a_2$, $a_3$ and $a_4$ are linear, quadratic and cubic functions of the superfield respectively. Since $a_2$ is linear in the parameters $\Sigma^{klmn}$ and the superfield, one cannot write any structure compatible with the helicity. Therefore, $a_2$ must be identically zero and the commutation relations (\ref{a3}) then imply $a_4=0$. In a similar manner, one can argue that all terms with even powers of the coupling constant vanish. \\\\
	
	\ndt We now arrive at a `universal form' for $\delta_{K}$ from the third commutator in (\ref {a3}). We find
	\bea
	\label{nl}
	\delta_{K}\,\phi\,&=&\,-  \, \frac{2}{\kappa}\theta^{\,klmn}_{\alpha\beta\gamma\delta}\;\Sigma_{klmn}\;+\;\xi(d)\,\frac{\kappa}{4!}\Sigma^{\,klmn}\frac{1}{{\del}^{+\,2}}\biggl (\,\mathcal{C}\left[\bar{d}^{\,\,\alpha\beta\gamma\delta}_{klmn}\,\frac{1}{\parp}\phi\,\del^{+\,3}\phi\right]\biggr)\,+\mathcal{O}(\kappa^3)\ ,\nn\\
	&&\;\;\;\nn\\
	&& \theta^{\,klmn}_{\alpha\beta\gamma\delta}\,\equiv \theta^{\,k\alpha}\,\theta^{\,l\beta}\,\theta^{\,m\gamma}\,\theta^{\,n\delta}\;\;,\;\; \bar{d}^{\,\alpha\beta\gamma\delta}_{klmn} \,\equiv\,\bar{d}_{\,k\alpha}\,\bar{d}_{\,l\beta}\,\bar{d}_{\,m\gamma}\,\bar{d}_{\,n\delta}\ ,
	\eea
	\ndt where $\xi(d)$ is a constant. The term $\mathcal{C}$ refers to the process of chiralization: the addition of terms (often polynomial in the superfields) to the existing superspace expression such that $d^{m\alpha}\,\delta_K\phi=0$~\cite{Ananth:2005vg}. This is essential to ensure that the operation preserves chirality, ie. the result of acting on a chiral expression is also chiral. \\

\ndt The table below summarizes the global symmetry, its compact subgroup (R-symmetry) and the little group for dimensions four through seven.
	
	\begin{longtable} [c] {| c | c | c | c|}
		\hline
		Dimension & Little group & R-symmetry & Global symmetry  \\
		\hline
		4   & SO(2) & SU(8) & $E_{7(7)}$ \\
		\hline
		5  & SO(3) &  USp(8) & $E_{6(6)}$ \\
		\hline
		6 & SO(4) & SO(5) $\times$ SO(5) & SO(5,5)\\
		\hline
		7 & SO(5) & SO(5) & SL(5,R)\\
		\hline
	\end{longtable}
	
	\ndt We now examine the construction of the symmetry generators in each of these dimensions to confirm the structure of the universal form in (\ref{nl}).
	
	\section{Evidence for the universal form}
	
	\subsection{$E_{7(7)}$ in four dimensions}
	 
	  \ndt In four dimensions, the adjoint of $E_{7(7)}$ splits up in terms of $SU(8)$ representations as $\mathbf{133}=\mathbf{63}+\mathbf{70}$. The $\mathbf{63}$ is linearly realized
	\bea
	\label{su8}
	\delta\phi\equiv\delta_{SU(8)}\phi~=~\omega^j_{\;k}\,\frac{1}{i\sqrt{2}\,\del^+}(q^k\,\bar{q}_j-\frac{1}{8}\,\delta^j_k\,q^m \bar{q}_m)\phi\,,
	\eea
	\ndt where the $\omega^j_{\;k}$ ($j,k=1,\ldots,8$) are 63 constant parameters. The $\mathbf{70}$ non-linearly realized coset generators are
	\bea
	\label{e7}
	&&\boldsymbol{\delta}\phi \equiv \delta_{E_{7(7)}/SU(8)}\phi=-\frac{2}{\kappa}\theta^{klmn}\overline{\Xi}_{klmn} \\
	&&+\;\frac{\kappa}{4!}\Xi^{klmn}\frac{1}{{\del}^{+\,2}}\biggl (\bar{d}_{klmn}\frac{1}{\parp}\phi\,\del^{+\,3}\phi-4\,\bar{d}_{klm}\,\phi\,\bar{d}_n\,\del^{+\,2
	}\phi+3\,\bar{d}_{kl}\,\parp\phi\,\bar{d}_{mn}\,\parp\phi\biggr)\ ,\nn
	\eea
	
	\ndt where $\Xi^{klmn}$ is the rank four completely anti-symmetric tensor corresponding to the 70 constant parameters and satisfies $\Xi^{klmn}=\frac{1}{2}\epsilon^{klmnpqrs}\,\overline{\Xi}_{pqrs}$. Here $\theta^{klmn}=\theta^k\theta^l\theta^m\theta^n$ and $\bar{d}_{i_1i_2...i_n}=\bar{d}_{i_1}\bar{d}_{i_2}...\bar{d}_{i_n}$.\\
	
	\ndt The $\delta\phi$ and the order $\kappa^{-1}$ piece in $\boldsymbol{\delta}\phi$ are chiral, trivially. To confirm `chirality' at the next order in (\ref{e7}), we act with the chiral derivative $d^{\,p}$ to obtain
	\bea
	d^p\boldsymbol{\delta}\phi~&=&~\frac{\kappa}{4!}\,\Xi^{klmn}\,\frac{1}{\del^{+\,2}}\biggl(d^p\,\bar{d}_{klmn}\frac{1}{\parp}\phi\,\del^{+\,3}\phi\,-\,4\,d^p\,\bar{d}_{klm}\,\phi\,\bar{d}_n\,\del^{+\,2}\phi\nn\\
	&&+\,4\,\bar{d}_{klm}\,\phi\,d^p\,\bar{d}_n\,\del^{+\,2}\phi\,+\,6\,d^p\,\bar{d}_{kl}\,\parp\phi\,\bar{d}_{mn}\,\parp\phi\biggr)\ ,
	\eea
	\ndt and use $\{d^m,\bar{d}_n\}~=~-i\sqrt{2}\,\delta^m_n\,\del^+$ to simplify this to
		\bea
	d^p\boldsymbol{\delta}\phi~&=&~-i\sqrt{2}\,\frac{\kappa}{4!}\,\Xi^{klmp}\,\frac{1}{\del^{+\,2}}\biggl(4\,\bar{d}_{klm}\phi\,\del^{+\,3}\phi\,-\,12\,\bar{d}_{kl}\del^+\,\phi\,\bar{d}_m\,\del^{+\,2}\phi\nn\\
	&&-\,4\,\bar{d}_{klm}\,\phi\,\del^{+\,3}\phi\,+\,12\,\bar{d}_{kl}\,\parp\phi\,\bar{d}_{m}\,\del^{+\,2}\phi\biggr)\,=\,0\ ,
	\eea
	
	\ndt confirming that (\ref{e7}) conforms to the universal form (\ref{nl}) described in section 3.

	\subsection{$E_{6(6)}$ in five dimensions}
	
	In five dimensions, the little group is $SO(3)\sim SU(2)$ and the role of $SU(8)$ is played by $USp(8)$~\cite{Cremmer:1979uq,Cremmer:1980gs}. We introduce symplectic Grassmann variables $\theta^{m\alpha}$ satisfying $\theta^{m\alpha}\,=\,C^{mn}\,\epsilon^{\alpha\beta}\,\bar{\theta}_{n\beta}$ where $m,n=1,2..8$, $\alpha,\beta=1,2$ and the $C_{mn}$ is the anti-symmetric symplectic invariant satisfying $C^{mn}C_{np}=-\delta^m_p$. The corresponding supercharges and covariant derivatives are denoted by $q^{m\alpha}$ ($\bar{q}_{m\alpha}$) and $d^{m\alpha}$($\bar{d}_{m\alpha}$) and satisfy
	\bea
	\label{algebra2}
	&&\{q^{m\alpha},\bar{q}_{n\beta}\}~=~i\sqrt{2}\,\delta^m_{\,n}\,\delta^{\alpha}_{\,\beta}\,\del^+\;\;\;\;;\;\;\;\;\{q^{m\alpha},q^{n\beta}\}~=~-i\sqrt{2}\,C^{mn}\,\epsilon^{\alpha\beta}\,\del^+\,,\nn\\
	&&\{d^{m\alpha},\bar{d}_{n\beta}\}~=~-i\sqrt{2}\,\delta^m_{\,n}\,\delta^{\alpha}_{\,\beta}\,\del^+\;\;\;\;;\;\;\;\;\{d^{m\alpha},d^{n\beta}\}~=~+i\sqrt{2}\,C^{mn}\,\epsilon^{\alpha\beta}\,\del^+.
	\eea
	\ndt The covariant derivatives anti-commute with the supercharges as usual. Since $\theta^{m2}=-C^{mn}\bar{\theta}_{n1}$, it suffices to work with the first component of the supercharges and covariant derivatives. With $\alpha=\beta=1$, the algebra of the supercharges and covariant derivatives is the same as in $SU(8)$ (\ref{su88}).\\

	\ndt The adjoint of $E_{6(6)}$ splits up in terms of $USp(8)$ representations as $\mathbf{78}=\mathbf{36}+\mathbf{42}$. The $\mathbf{36}$ is
	\bea
	\label{t}
	\delta\phi=w_{mn}T^{mn}\,\phi\;\;; T^{mn}=\frac{1}{i\sqrt{2}\parp}(q^m\,C^{np}\,\bar{q}_p\,+\,q^n\,C^{mp}\,\bar{q}_p)\ ,
	\eea
	where $q^{m}\equiv q^{m1}$ , $\bar{q}_{m}\equiv\bar{q}_{m1}$ and $w_{mn}$ are the 36 parameters. The $T^{mn}$ satisfy the algebra
	\bea
	[T^{kl},T^{mn}]~=~C^{lm}\,T^{kn}+C^{ln}\,T^{km}+C^{km}\,T^{ln}+C^{kn}\,T^{lm}\ .
	\eea
	\ndt The remaining $\mathbf{42}$ constitute non-linear transformations given by
	\bea
	\label{e6}
	&&\delta_{E_{6(6)}/USp(8)}\phi~=~-\frac{2}{\kappa}\theta^{klmn}\Sigma_{klmn} \\
	&&+\;\frac{\kappa}{4!}\Sigma^{klmn}\frac{1}{{\del}^{+\,2}}\biggl (\bar{d}_{klmn}\frac{1}{\parp}\phi\,\del^{+\,3}\phi-4\,\bar{d}_{klm}\,\phi\,\bar{d}_n\,\del^{+\,2
	}\phi+3\,\bar{d}_{kl}\,\parp\phi\,\bar{d}_{mn}\,\parp\phi\biggr)\ ,\nn
	\eea
	\ndt
	with $\theta^{klmn}\equiv\theta^{k1}\theta^{l1}\theta^{m1}\theta^{n1}$ and $\bar{d}_{klmn}\equiv\bar{d}_{k1}\bar{d}_{l1}\bar{d}_{m1}\bar{d}_{n1}$. The $\Sigma^{klmn}$ is the rank 4 completely anti-symmetric traceless tensor  and satisfies  $(\Sigma^{klmn})^{*}=\Sigma_{klmn}$. Notice that the chirality constraint is satisfied by (\ref{e6}) for the same `structural' reasons as in (\ref{e7}). This confirms that the transformations in (\ref{e6}) are also of the form in (\ref{nl}). \\
	
\vskip 0.3cm
	 
	 \subsection{$E_{5(5)}$ in six dimensions}
	 
	 \ndt Six dimensions is the first dimension where the global symmetry of the maximal supergravity theory is a classical group: $SO(5,5)$. This group has the compact subgroup $SO(5)\times SO(5)$, which is the R-symmetry of the theory. The little group is $SO(4)\sim SU(2)\times SU(2)$. We use the Lie algebra isomorphism $SO(5)\sim USp(4)$ to define 
	 the Grassmann variables $(\theta^{m\alpha},\theta^{a\dot{\alpha}})$ with $m,a=1,2,..4$ being the internal symmetry indices while $\alpha,\dot{\alpha}=1,2$ represent little group indices. We denote the $USp(4)$ symplectic invariants by $C_{mn}$ and $C_{ab}$. The supercharges $(q^{m\alpha},q^{a\dot{\alpha}})$ and the covariant derivatives $(d^{m\alpha},d^{a\dot{\alpha}})$  within each $SU(2)$ satisfy the algebra (\ref{algebra2}), while they anti-commute with those from the other $SU(2)$. As before (line below equation (\ref {algebra2})), we can work with $\alpha=1,\dot{\alpha}=1$. \\
	  
\ndt The $SO(5)$ gamma matrices $\gamma^I$ ($I=1,2..5$) satisfy
\bea
\label{gamma}
\{\gamma^{I}\,,\gamma^{J}\,\}~=~2\,\delta^{IJ}\,\mathbb{I}_{4\times4}\ ,
\eea  
\ndt where the $(\gamma^I)^{kl}=C^{lm}(\gamma^I)^{k}_{\,\,\,m}$ are anti-symmetric in $k,l$ and traceless with respect to $C_{kl}$~\cite{Brink:2010ti}. 
\ndt A vector of $SO(5)$ can then be written as a bi-spinor 
\bea
p_{mn}~=~p_I\,(\gamma^{I})_{mn}\,.
\eea
\ndt The commutators of the gamma matrices $(\gamma^{IJ})^{kl}=C^{lm}(\gamma^{IJ})^{k}_{\,\,\,m}$ are symmetric in $k,l$. \\

	 \ndt The adjoint of $SO(5,5)$ splits up in terms of $SO(5)\times SO(5)$ representations as $\mathbf{45}=\mathbf{20}+\mathbf{25}$. The $\mathbf{20}$ linearly realized generators of $SO(5) \times SO(5)$ are spanned by $T^{IJ}$ and $T^{AB}$ ($I,J$ and $A,B$ correspond to the two independent $SO(5)$ groups)
	 \bea
	 \label{t}
	 &&T^{IJ}~=~(\gamma^{IJ})_{mn}\,\frac{1}{i\sqrt{2}\parp}(q^m\,C^{nk}\,\bar{q}_k\,+\,q^n\,C^{mk}\,\bar{q}_k)\,,\nn\\
	 &&T^{AB}~=~(\gamma^{AB})_{ab}\,\frac{1}{i\sqrt{2}\parp}(q^a\,C^{bc}\,\bar{q}_c\,+\,q^b\,C^{ac}\,\bar{q}_c)\,,
	 \eea
	 	\ndt where  $q^{m}\equiv q^{m1}$ , $\bar{q}_{m}\equiv\bar{q}_{m1}$ and  $q^{a}\equiv q^{a1}$ , $\bar{q}_{a}\equiv\bar{q}_{a1}$. \\
	 	
	  \ndt The $\mathbf{25}$ non-linear transformations are given by
	\bea
	\label{e5}
	&&\delta_{E_{5(5)}/H}\phi~=~-\frac{2}{\kappa}\theta^{klab}\Sigma_{klab}\;+\;\kappa\,\Sigma^{klab}\frac{1}{{\del}^{+\,2}}\left(\mathcal{C}\biggl [\bar{d}_{klab}\frac{1}{\parp}\phi\,\del^{+\,3}\phi\biggr]\right)\ ,\nn
	\eea
	  
	  \ndt where $\Sigma_{klab}=\Sigma_{IA}\,(\gamma^{I})_{kl}\,(\gamma^{A})_{ab}$, $\theta^{klab}\equiv\theta^{k1}\theta^{l1}\theta^{a1}\theta^{b1}$ and $\bar{d}_{klmn}\equiv\bar{d}_{k1}\bar{d}_{l1}\bar{d}_{a1}\bar{d}_{b1}$. The $\Sigma_{IA}$ are the 25 constant parameters. The chiral function is structurally different (from the one in four and five dimensions) and reads
	  \bea
	  \label{chiral}
	  \mathcal{C}\biggl [\bar{d}_{klab}\frac{1}{\parp}\phi\,\del^{+\,3}\phi\biggr]~&=&~\bar{d}_{klab}\frac{1}{\parp}\phi\,\del^{+\,3}\phi\;-2\,\bar{d}_{kla}\,\phi\,\bar{d}_b\,\del^{+\,2}\phi\nn\\
	  &&\hspace{-4mm}-2\,\bar{d}_{kab}\,\phi\,\bar{d}_l\,\del^{+\,2}\phi\;+\,\bar{d}_{kl}\,\parp\phi\,\bar{d}_{ab}\,\parp\phi\,+2\,\bar{d}_{ka}\,\parp\phi\,\bar{d}_{bl}\,\parp\phi.
	  \eea
	 \ndt We confirm that $\delta_{E_{5,5}/H}\phi\equiv\delta\phi$ is chiral by acting with $d^m\equiv d^{m1}$
	 \bea
	 d^m\delta\phi~&=&~\kappa\,\Sigma^{klab}\,\biggl(d^m\,\bar{d}_{klab}\frac{1}{\parp}\phi\,\del^{+\,3}\phi\,-2\,d^m\,\bar{d}_{kla}\,\phi\,\bar{d}_b\,\del^{+\,2}\phi\,-\,2\,d^m\,\bar{d}_{kab}\,\phi\,\bar{d}_l\,\del^{+\,2}\phi\nn\\
	 &&+2\,\bar{d}_{kab}\,\phi\,d^m\,\bar{d}_l\,\del^{+\,2}\phi\,+\,d^m\,\bar{d}_{kl}\,\parp\phi\,\bar{d}_{ab}\,\parp\phi\,+\,4\,d^m\,\bar{d}_{ka}\,\parp\phi\,\bar{d}_{bl}\,\parp\phi\biggr)\ ,
	 \eea
	  \ndt which simplifies to
	  \bea
	  d^m\delta\phi~&=&~-i\sqrt{2}\,\kappa\,\Sigma^{mlab}\,\biggl(2\,\bar{d}_{lab}\,\phi\,\del^{+\,3}\phi\,-4\,\bar{d}_{la}\,\del^+\,\phi\,\bar{d}_b\,\del^{+\,2}\phi\,-2\,\bar{d}_{ab}\,\del^+\,\phi\,\bar{d}_l\,\del^{+\,2}\phi\nn\\
	  &&-\,2\,\bar{d}_{lab}\,\phi\,\del^{+\,3}\phi\,+2\,\bar{d}_{ab}\,\del^+\,\phi\,\bar{d}_l\,\del^{+\,2}\phi+4\,\bar{d}_{la}\,\del^+\,\phi\,\bar{d}_b\,\del^{+\,2}\phi\,\biggl)\,=\,0\,.
	  \eea
	 \ndt It can be similarly shown by acting with $d^a\equiv d^{a1}$ that $d^a\delta\phi=0$. Hence, the transformations (\ref{e5}) are of the form in (\ref{nl}).

\vskip 0.3cm

	\subsection{$E_{4(4)}$ in seven dimensions}

	 \ndt In seven dimensions, the global symmetry $SL(5,R)$ has the compact subgroup $SO(5)$, which is also the massless little group. We use the Lie algebra isomorphism $SO(5)\sim USp(4)$ to define the Grassmann variables $\theta^{m\alpha}$ satisfying $\theta^{m\alpha}=C^{mn}C^{\alpha\beta}\bar{\theta}_{n\beta}$, where both $m,\alpha$ are $USp(4)$ indices. The corresponding supercharges and covariant derivatives satisfy the algebra
	 	\bea
	 \label{algebra3}
	 &&\{q^{m\alpha},\bar{q}_{n\beta}\}~=~i\sqrt{2}\,\delta^m_{\,n}\,\delta^{\alpha}_{\,\beta}\,\del^+\;\;\;\;;\;\;\;\;\{q^{m\alpha},q^{n\beta}\}~=~-i\sqrt{2}\,C^{mn}\,C^{\alpha\beta}\,\del^+\,,\nn\\
	 &&\{d^{m\alpha},\bar{d}_{n\beta}\}~=~-i\sqrt{2}\,\delta^m_{\,n}\,\delta^{\alpha}_{\,\beta}\,\del^+\;\;\;\;;\;\;\;\;\{d^{m\alpha},d^{n\beta}\}~=~+i\sqrt{2}\,C^{mn}\,C^{\alpha\beta}\,\del^+.
	 \eea
	 \ndt We choose a form for the symplectic invariant $C_{\alpha\beta}$~\cite{Brink:2010ti}
	 \bea
	 C_{\alpha\beta}= \begin{pmatrix}
	 	0 & 0 & 0 & 1 \\
	 	0 & 0 & 1 & 0 \\
	 	0 & -1 & 0 & 0 \\
	 	-1 & 0 & 0 & 0
	 \end{pmatrix}
	 \eea
	 \ndt It is then easy to check that with $\alpha,\beta=1,2$ in (\ref{algebra3}), the algebra of supercharges and covariant derivatives is similar to (\ref{su88}). We work with these components to write the $SL(5,R)$ transformations.\\
	 
	\ndt The adjoint of $SL(5,R)$ splits up in terms of $SO(5)$ representations as $\mathbf{24}=\mathbf{10}+\mathbf{14}$. We use the $SO(5)$ gamma matrices defined in (\ref{gamma}) to write the $\mathbf{10}$ linear transformations $T^{IJ}$ ($I,J=1,2,..5$)
	\bea
	T^{IJ}~=~(\gamma^{IJ})_{mn}\,\frac{1}{i\sqrt{2}\,\parp}(q^{m1}\,C^{nk}\,\bar{q}_{k1}\,+\,q^{n1}\,C^{mk}\,\bar{q}_{k1}\,+\,q^{m2}\,C^{nk}\,\bar{q}_{k2}\,+\,q^{n2}\,C^{mk}\,\bar{q}_{k2})\,.
	\eea
	
	\ndt The $\mathbf{14}$ non-linear transformations are given by
	 \bea
	 \label{e4}
	 &&\delta_{SL(5,R)/SO(5)}\phi~=~-\frac{2}{\kappa}\theta^{kl\underline{m}\underline{n}}\Sigma_{klmn}\;+\;\kappa\,\Sigma^{klmn}\frac{1}{{\del}^{+\,2}}\left(\mathcal{C}\biggl [\bar{d}_{kl\underline{m}\underline{n}}\frac{1}{\parp}\phi\,\del^{+\,3}\phi\biggr]\right) ,
	 \eea
	  
	\ndt where $\theta^{kl\underline{m}\underline{n}}\equiv \theta^{k1}\,\theta^{l1}\,\theta^{m2}\,\theta^{n2}$ and $\bar{d}_{kl\underline{m}\underline{n}}\equiv\bar{d}_{k1}\,\bar{d}_{l1}\,\bar{d}_{m2}\,\bar{d}_{n2}$. The chiral function in (\ref{e4}) is the same as that in (\ref{chiral}). $\Sigma_{klmn}\,=\,\Sigma_{IJ}\,(\gamma^{I})_{kl}\,(\gamma^{J})_{mn}$ and the $\Sigma_{IJ}$ is a symmetric traceless tensor representing the $14$ constant parameters.\\
	
	\ndt Thus the global symmetries in dimensions four through seven share this universal form. This approach may also be useful to understand how to formulate the eleven-dimensional model in a manner where these global symmetries are made explicit~\cite{Hohm:2013pua,Hohm:2019bba}. In principle, the global symmetries in eight and nine dimensions, can also be constructed in a similar fashion. It would also be interesting to perform a similar analysis of symmetries in the $d<4$ models which include the infinite dimensional Kac-Moody algebras~\cite{West:2001as,Damour:2002et,Damour:2002cu,Henneaux:2007ej,Bossard:2018utw}.\\
	
	\section{An aside: the Lagrangians}
	
	\ndt While not the main focus of this paper, it is interesting to look at the actions of the theories discussed so far. Each of the actions may be obtained by starting from the $d=4$ action and `oxidizing' the theory by introducing a `generalized derivative'~\cite{Ananth:2005vg}. Specifically, the Lagrangian for the five dimensional theory was obtained in this manner in~\cite{Ananth:2024liy}: first, the superfield $\phi$ is given added dependance on the additional coordinate $x^5$, then the kinetic term is obtained by including the extra transverse derivative in the d'Alembertian of (\ref{La}) and finally, the cubic vertex is obtained by replacing $\bar{\del}$ with $\overline{\nabla}$ (and similarly for its complex conjugate) where
	\bea
	\overline{\nabla}~=~\bar{\del}\,+\,\frac{\sigma}{16}\,C^{ij}\,\bar{d}_i\,\bar{d}_j\,\frac{\del^5}{\del^+}\,,
	\eea 
	\ndt and $C_{ij}$ is the $USp(8)$ symplectic invariant. We check that this generalized derivative has appropriate transformations under the little group $SO(3)$ (appendix A.1 lists the relevant generators). We perform two rotations 
\bea
[\,\overline{\nabla}\,,\,j^5\,]~=~\nabla^5~=~-i\,\del^5\,+\,i\,\frac{\sigma}{16}\,\bar{d}_m\,C^{mn}\,\bar{d}_{n}\,\frac{\del}{\parp}\ ,
\eea
and
\bea
[\,\nabla^5\,,\,\bar{j}^5\,]~=~\overline{\nabla}\ ,
\eea
\ndt thereby proving that $(\overline{\nabla},\nabla^5)$ transforms as a vector under the little group $SO(3)$. The exact value of the constant $\sigma$ is fixed by demanding Lorentz invariance of the resulting cubic vertex.

\vskip 0.3cm 
	\ndt In a similar manner, the Lagrangians in six and seven dimensions may be constructed, starting from the $d=4$ model, using the generalized derivatives
	
	\bea
	\overline{\nabla}^{(6)}~=~\bar{\del}\,+\,\frac{\eta}{16}\,C^{ij}\,\bar{d}_i\,\bar{d}_j\,\frac{\del^5}{\del^+}\,+\,\frac{\eta}{16}\,C^{ab}\,\bar{d}_a\,\bar{d}_b\,\frac{\del^6}{\del^+}	\,,
	\eea
		
		\bea
		 \overline{\nabla}^{(7)}~=~\bar{\del}\,+\,\frac{\chi}{16}\,C^{ij}\,\bar{d}_i\,\bar{d}_j\,\frac{\del^5}{\del^+}\,+\,\frac{\chi}{16}\,C^{ij}\,\bar{d}_{\underline{i}}\,\bar{d}_{\underline{j}}\,\frac{\del^6}{\del^+}\,\
		+\,\frac{\chi}{16}\,C^{ij}\,\bar{d}_{\underline{i}}\,\bar{d}_j\,\frac{\del^7}{\del^+}\, .
		\eea
		
	\ndt The constants $\eta$ and $\chi$ are again determined using the requirement of Lorentz invariance.

	\begin{center}
	* ~ * ~ *
\end{center}

\ndt  {\bf{Towards an all-order expression for $E_7\,$:}}  an obvious limitation of our approach is that it is perturbative - and hence needs new derivations at each new order in the coupling constant. This makes it important to examine how higher order or even all-order results may be arrived at in this formalism. In this section, we discuss this in the context of $E_7$ in $d=4$. Any results obtained in this manner may be extended to higher dimensions since we have already shown how global symmetries in $d>4$ theories mimic the structure of the $E_7$ in four dimensions.\\

\ndt As mentioned already, the $SU(8)$ transformations are linear while the coset $E_{7(7)}/SU(8)$ generators are a series in odd powers of the coupling constant. These non-linear transformations, at higher orders, may be found from the requirement of closure (of the commutator of two such transformations  on $SU(8)$). We demonstrate this specifically for order $\kappa^3$ by writing all the relevant contributions
\bea
\label{const}
\{[\,\boldsymbol{\delta}_1^{\kappa^{-1}}\,,\,\boldsymbol{\delta}_2^{\kappa^3}\,]\,+\,[\,\boldsymbol{\delta}_1^{\kappa^3}\,,\,\boldsymbol{\delta}_2^{\kappa^{-1}}\,]\,+\,[\,\boldsymbol{\delta}_1^{\kappa}\,,\,\boldsymbol{\delta}_2^{\kappa}\,]\}\phi\,=\,0.
\eea

\ndt where $\boldsymbol{\delta}$ denotes the coset variation. The third term in the above equation is given by 
\bea
\label{k2}
\kappa^2\,(\Xi^{klmn}_{1}\,\Xi^{pqrs}_{2}\,-\,\Xi^{klmn}_{2}\,\Xi^{pqrs}_{1})\,\phi_{klmnpqrs},
\eea
where
\bea
\phi_{klmnpqrs}\,&=&\,\frac{1}{4!}\frac{1}{\del^{+\,2}}\biggl (\bar{d}_{klmn}\frac{1}{\parp}\phi_{pqrs}\,\del^{+\,3}\phi\,+\,\bar{d}_{klmn}\frac{1}{\parp}\phi\,\del^{+\,3}\phi_{pqrs}\nn\\
&&-4\,\bar{d}_{klm}\,\phi_{pqrs}\,\bar{d}_n\,\del^{+\,2
}\phi\,-\,4\,\bar{d}_{klm}\,\phi\,\bar{d}_n\,\del^{+\,2
}\phi_{pqrs}\,\nn\\
&&+3\,\bar{d}_{kl}\,\parp\phi_{pqrs}\,\bar{d}_{mn}\,\parp\phi\,+\,3\,\bar{d}_{kl}\,\parp\phi\,\bar{d}_{mn}\,\parp\phi_{pqrs}\biggr).
\eea
\ndt The function $\phi_{pqrs}$ is chiral and is given by
\bea
\phi_{pqrs}\,=\,\frac{1}{4!}\frac{1}{\del^{+\,2}}\biggl (\bar{d}_{pqrs}\frac{1}{\parp}\phi\,\del^{+\,3}\phi-4\,\bar{d}_{pqr}\,\phi\,\bar{d}_s\,\del^{+\,2
}\phi+3\,\bar{d}_{pq}\,\parp\phi\,\bar{d}_{rs}\,\parp\phi\biggr)\ ,
\eea

\ndt implying that the function $\phi_{pqrsklmn}$ is also chiral. We conjecture the following form at $\kappa^3$
\bea
\label{k3}
\boldsymbol{\delta}^{\kappa^3}\phi\,=\,\kappa^3\,\Xi^{klmn}\,\epsilon^{pqrsabcd}\,\phi_{klmnpqrs}\,\bar{d}_{abcd}\phi + \rho(\phi).
\eea

\ndt We substitute this expression in (\ref{const}), to find that when the $\boldsymbol{\delta}^{\kappa^{-1}}$ acts on the last superfield of the first term in (\ref{k3}), it cancels piece (\ref{k2}). $\boldsymbol{\delta}^{\kappa^{-1}}\,\phi_{klmnpqrs}\ne 0$ allows us to fix $\rho(\phi)$ from equation (\ref{const}). One interesting finding here is that the chiral function that appears in (\ref{k3}) is the same as in (\ref{e7}), in a nested form. This suggests that it may be possible to write an all order expression for the $E_7$ transformations in a compact manner. Since these transformations relate interaction vertices at different orders in the coupling constant~\cite{Ananth:2024liy}, they will play an important role in determining the ultra-violet properties of the theory. 

\vskip 0.3cm

\section*{Acknowledgments}

\ndt NB acknowledges the SRF-NET fellowship by the Council of Scientific and Industrial Research (CSIR), India.
\newpage
 \appendix
 
 \section{The $d=4$ superPoincar\'e algebra}
	
	\ndt On the light-cone, the SuperPoincar\'e algebra splits up into kinematical (independent of interactions) and dynamical generators (dependent on interactions)~\cite{Bengtsson:1983pg, Ananth:2005vg}.
	\subsection*{Bosonic generators}
	 The kinematical generators at the light-cone time $x^+=0$ are comprised of the momenta
	\be
	p^+_{}~=~-i\,\partial^+_{}\ ;\qquad p~=~-i\,\partial\ ; \qquad \bar p~=~-i\,\bar\partial\ ,
	\ee
	\newline
	\ndt the rotation in the transverse plane
	\be
	j~=~x\,\bar\partial-\bar x\,\partial+ S^{12}_{}\ ,
	\ee
	where
	\be
	S^{12}_{}~=~ \frac{1}{4i\sqrt{2}\parp}(q^j\bar{q}_j-\bar{q}_jq^j),
	\ee
	\ndt and the rotations
	\be
	j^+_{}~=~i\, x\,\partial^+_{}\ ;\qquad   \bar j^+_{}~=~i\,\bar x\,\partial^+_{}\ ,
	\ee
	\be
	j^{+-}_{}~=~i\,x^-_{}\,\partial^+_{}-\frac{i}{2}\,(\,\theta^m_{}\frac{\del}{\del\bar{\theta}_m}+\bar\theta^{}_m\,\frac{\del}{\del\theta^m}\,)\ .
	\ee
	\ndt The dynamical generators include the light-cone Hamiltonian
	\be
	p^-_{}~=~-i\frac{\partial\bar\partial}{\partial^+_{}}
	\ee
	\newline
	 and the boosts
	\bea
	\label{boost}
	j^-_{}&=&i\,x\,\frac{\partial\bar\partial}{\partial^+_{}} ~-~i\,x^-_{}\,\partial~+~i\,\Big( \theta^m_{}\frac{\del}{\partial\theta^m}\,+\frac{i}{4\sqrt{2}\,\partial^+}\,(\,d^m\,\bar d_m-\bar d_m\,d^m\,)\Big)\frac{\partial}{\partial^+_{}}\,\ ,\cr 
	\bar j^-_{}&=&i\,\bar x\,\frac{\partial\bar\partial}{\partial^+_{}}~ -~i\,x^-_{}\,\bar\partial~+~ i\,\Big(\bar\theta_m^{}\frac{\del}{\partial\bar{\theta}_m}+\frac{i}{4\sqrt{2}\,\partial^+}\,(\,d^m\,\bar d_m-\bar d_m\,d^m\,)\,\Big)\frac{\bar\partial}{\partial^+_{}}\,\ .
	\eea

	\subsection*{Fermionic generators}
	
	\vskip 0.3cm
	\ndt The kinematical supersymmetries are given by
	\be
	\label{su8ks}
	q^{\,m}=-\frac{\del}{\del\bar{\theta}_m}\,+\,\frac{i}{\sqrt 2}\,{\theta^m}\,{\partial^+}\ ;\qquad{{\bar q}_{\,m}}=\;\;\;\frac{\del}{\del\theta^m}\,-\,\frac{i}{\sqrt 2}\,{{\bar \theta}_m}\,{\partial^+}\ ,
	\ee
	while the dynamical supersymmetries read
	\bea
	\label{dynsus}
	\begin{split}
		Q^m\,\equiv\,&i\,[\,\bar j^-\,,\,q^m\,]\,=\,\frac{\bar \partial}{\parp}\,q^m\,  , \\
		\bar{Q}_n\,\equiv\,&i\,[\,j^-\,,\,{\bar q}_{\,n}\,]\,=\,\frac{\partial}{\parp}\,{\bar q}_{\,n}\, .
	\end{split}
	\eea
	
	\subsection{Upgrading the superPoincar\'e algebra to $d=5$}
	\ndt In five dimensions, the little group $SO(2)$ in four  dimensions is upgraded to $SO(3)$. We therefore include the coset generators $SO(3)/SO(2)$
	\bea
	j^5~&=&~i\,(\,x^5\,\del\,-\,x\,\del^5\,)\,-\,\frac{1}{4\,i\,\del^+}\,q^k\,C_{kl}\,q^l\ ,\nn\\
	\bar{j}^5~&=&~i\,(\,x^5\,\bar{\del}\,-\,\bar{x}\,\del^5\,)\,+\,\frac{1}{4\,i\,\del^+}\,\bar{q}_k\,C^{kl}\,\bar{q}_l\,.
	\eea
	\ndt The $SO(3)$ algebra is spanned by $j\,,\,j^{5}\,,\,\bar{j}^{5}$ which obey
	\bea
	[\,j^5\,,\,\bar{j}^5\,]~=~j\;;\;\;\;\;\;\;\;\;[\,j\,,\,j^5\,]~=+j^5\;;\;\;\;\;\;\;\;[\,j\,,\,\bar{j}^5\,]~=-\bar{j}^5.
	\eea
	Additional generators include the rotation $j^{+5}\,=\,i\,x^5\,\del^+$ (at $x^+=0$) and the momenta $p^5\,=\,-i\,\del^5$. The expression for $p^-$ now reads
	\be
	p^-_{}~=~-i\frac{\partial\bar\partial+\frac{1}{2}\del^5\del^5}{\partial^+_{}}.
	\ee
	 The first term in the expressions for $j^-$ and $\bar{j}^-$ (\ref{boost}) is similarly modified . The dynamical generator $j^{-5}$ is obtained from the commutator $[\,j^-\,,\,\bar{j}^5\,]\,=\,j^{-5}$. \\
	 
	 \ndt Similarly the generators corresponding to the superPoincar\'e groups in six and seven dimensions can be constructed.

	 \newpage
	 \section{How the degrees of freedom re-shuffle in higher-dimensions}
	 
	 \ndt In our analysis, we use the same four dimensional superfield $\phi$ in each higher dimension since all the propagating degrees of freedom are contained in it. We explain in this section, how the four dimensional degrees of freedom are re-packaged inside the superfield in higher dimensions using the relevant symmetry groups.\\
	 
	 \ndt In five dimensions, the little group is upgraded from $SO(2)$ to $SO(3)$. All the physical fields are then the representations of $SO(3)$ labeled by half-integers $j$ with dimension $2j+1$. The R-symmetry $SU(8)$ is reduced to the $USp(8)$ group, ie. $SU(8)$ representations in the superfield split in terms of $USp(8)$ representations when expanded in terms of the Grassmann coordinates $\theta^{m1}$ where $m=1,2,...8$ is the $USp(8)$ index.\\
	 
	 \ndt  Specifically, the $\bf{70}$ of $SU(8)$ splits as
	 \bea
	 \bf{70}=\bf{42}+\bf{27}+\bf{1}.
	 \eea
	 \ndt We therefore get $\bf{42}$ scalar fields, $\bf{27}$ vector bosons and $\bf{1}$ graviton degree of freedom (each with Cartan eigenvalue zero). The $\bf{28}$ of $SU(8)$ splits into 
	 \bea
	 \bf{28}~=~\bf{27}+\bf{1}\ ,
	 \eea
	 \ndt producing $\bf{27}$ vector bosons and $\bf{1}$ graviton each with Cartan eigenvalue $+1$ (and $-1$ for the $\overline{\bf{28}}$ of $SU(8)$). The $\bf{56}$ of $SU(8)$ is now
	 \bea
	 \bf{56}~=~\bf{48}+\bf{8}\ ,
	 \eea
	 \ndt representing $\bf{48}$ $\text{spin-}1/2$ and $\bf{8}$ spin-$3/2$ fields with the Cartan eigenvalue $+1/2$ (and $-1/2$ for the $\overline{\bf{56}}$). \\
	 
	 \ndt The table below summarizes the degrees of freedom in five dimensions.
	 
	 \begin{longtable} [c] {| c | c | c | c |}
	 	\hline
	 	Degrees of freedom & $SO(3)$ representation & $USp(8)$ representation & Total\\
	 	  &  $A$ & $B$ &  $A\times B$\\
	 	\hline
	 	Gravitons & 5 & 1 & 5\\
	 	\hline
	 	Gravitinos & 4 & 8 & 32\\
	 	\hline
	 	Vectors & 3 & 27 & 81\\
	 	\hline
	 	Gauginos & 2 & 48 & 96\\
	 	\hline
	 	Scalars & 1 & 42 & 42\\
	 	\hline
	 \end{longtable}

	 \ndt In six dimensions, the physical fields are representations of the little group $SO(4)\sim SU(2)\times SU(2)$, labeled by two half-integers $(j_1,j_2)$ (corresponding to the two Cartan generators) with dimension $(2\,j_1\,+\,1)\times (2\,j_2\,+\,1)$. The R-symmetry is reduced to the $USp(4)\times USp(4)$ group. The superfield may be expanded in terms of the Grassmann coordinates $(\theta^{m1},\theta^{a1})$ where $m,a=1,2,..4$ correspond to the two $USp(4)$ groups. The $\mathbf{70}$ of $SU(8)$ is then
	 \bea
	 \mathbf{70}~=~\mathbf{5}\times \mathbf{5}\,+\,\mathbf{4}\times\mathbf{4}\,+\,\mathbf{4}\times\mathbf{4}\,+\,\mathbf{5}\times\mathbf{1}\,+\mathbf{1}\times\mathbf{5}\,+\mathbf{1}\,+\mathbf{1}\,+\mathbf{1}\,,
	 \eea
	 \ndt representing the 25 scalars, 32 vectors (with Cartan eigenvalues ($\pm\frac{1}{2}$\,,\,$\mp\frac{1}{2}$)), 10 anti-symmetric bosons and 1 graviton each with Cartan eigenvalue (0,0). We also get two gravitons with Cartan eigenvalues $(\pm1\,,\,\mp1)$. The $\mathbf{28}$ splits as
	 \bea
	 \mathbf{28}~=~\mathbf{4}\times \mathbf{4}\,+\,\mathbf{5}\times \mathbf{1}\,+\,\mathbf{1}\times \mathbf{5}\,+\,\mathbf{1}\,+\,\mathbf{1}\,,
	 \eea
	 \ndt producing 16 vectors with Cartan eigenvalue ($\frac{1}{2},\frac{1}{2}$), 5 anti-symmetric bosons and a graviton with Cartan eigenvalue (1,0) (and similarly with (0,1)). The $\overline{\mathbf{28}}$ yields the conjugate fields with the opposite helicity. The $\mathbf{56}$ is now
	 \bea
	 \mathbf{56}~=~\mathbf{5}\times \mathbf{4}\,+\,\mathbf{4}\times \mathbf{5}\,+\,\mathbf{1}\times \mathbf{4}\,+\,\mathbf{4}\times \mathbf{1}\,+\,\mathbf{1}\times \mathbf{4}\,+\,\mathbf{4}\times \mathbf{1}\,,
	 \eea
	\ndt producing the 20 gauginos and 4 gravitinos with Cartan eigenvalues $(\,\frac{1}{2}\,,\,0\,)$ (and similarly with $(\,0,\frac{1}{2}\,)$). We also get 4 gravitinos each with Cartan eigenvalue $(\,-\frac{1}{2}\,,\,1\,)$ (and similarly with $(\,1,-\frac{1}{2}\,)$). The $\overline{\mathbf{56}}$ splits in a similar manner yielding the conjugate fields (of opposite helicity). The table below summarizes the degrees of freedom in six dimensions~\cite{Schwarz:1980gf}.
	\begin{longtable} [c] {| c | c | c | c |}
	 	\hline
	 	Degrees of freedom & $SU(2)\times SU(2)$ & $USp(4)\times USp(4)$ & Total\\
	 	  &  $A$ & $B$ &  $A\times B$\\
	 	\hline
	 	Gravitons & 9 & 1 & 9\\
	 	\hline
	 	Gravitinos & 12 & 4 & 48\\
	 	\hline
	 	Vectors & 4 & 16 & 64\\
	 	\hline
	 	 Anti-symmetric bosons & 6 & 5 & 30\\
	 	\hline
	 	Gauginos & 4 & 20 & 80\\
	 	\hline
	 	Scalars & 1 & 25 & 25\\
	 	\hline
	 \end{longtable}
	 
	 \ndt This analysis can be extended to seven and higher dimensions similarly.

\end{document}